\documentclass[11pt]{article}

\usepackage{hyperref,amsfonts,amsmath,amssymb,graphicx,bm,subfigure,a4wide,cite}

\numberwithin{equation}{section}

\begin{document}

\title{\textbf{Gravitational scattering of spinning neutrinos by a rotating black
hole with a slim magnetized accretion disk}}

\author{Maxim Dvornikov\thanks{maxdvo@izmiran.ru}
\\
\small{\ Pushkov Institute of Terrestrial Magnetism, Ionosphere} \\
\small{and Radiowave Propagation (IZMIRAN),} \\
\small{108840 Moscow, Troitsk, Russia}}

\date{}

\maketitle

\begin{abstract}
We study neutrinos gravitationally scattered off a rotating supermassive
black hole which is surrounded by a thin accretion disk with a realistic
magnetic field. Neutrinos are supposed to be Dirac particles having
a nonzero magnetic moment. Neutrinos move
along arbitrary trajectories, with the incoming flux being parallel to the equatorial plane.
We exactly account for the influence of both gravity and the magnetic
field on the neutrino motion and its spin evolution. The general statement that the helicity of an ultrarelativistic neutrino is constant in the particle scattering in an arbitrary gravitational field is proven within the quasiclassical approach.
We find the measurable
fluxes of outgoing neutrinos taking into account the neutrino spin
precession in the external field in curved spacetime. These fluxes
turn out to be significantly suppressed for some parameters of the
system. Finally, we discuss the possibility to observe the predicted
phenomena for core-collapsing supernova neutrinos in our Galaxy.
\end{abstract}

\section{Introduction}

Neutrinos are experimentally confirmed to be massive and mixed particles
(see, e.g., Ref.~\cite{Ace21}). It results in transitions between
different neutrino types named neutrino flavor oscillations. Standard
model neutrinos are left particles, i.e. their spin is opposite to
the particle momentum. However, the neutrino polarization can change
under the influence of an external field. This process is called neutrino
spin oscillations. The combination of these two phenomena is also
possible. In this situation, we deal with neutrino spin-flavor oscillations.

External fields, i.e., the neutrino interaction with matter~\cite{Smi05}
and with an electromagnetic field~\cite{Giu16}, are known to affect
neutrino oscillations. The gravitational interaction, despite it
is quite weak, can also induce neutrino oscillations. Neutrino flavor~\cite{CarFul97},
spin~\cite{Dvo06}, and spin-flavor~\cite{PirRoyWud96} oscillations
in a gravitational field were previously studied. In the present work,
we deal with the evolution of a neutrino spin in a curved spacetime
under the influence of a magnetic field.

For the first time, the behavior of a spinning particle in a curved
spacetime was studied in Ref.~\cite{Pap51}. The dynamics of the
fermion spin in a gravitational field was analyzed in Ref.~\cite{ObuSilTer17}
basing on the Dirac equation in a curved spacetime. The quasiclassical
equation for a particle spin in a gravitational field was derived
in Ref.~\cite{PomKhr98}. The method of Ref.~\cite{PomKhr98} was
applied in Ref.~\cite{Dvo06} to describe neutrino oscillations in the vicinity of a nonrotating black hole (BH) in frames of the General Relativity (GR). Neutrino spin oscillations in various extensions of GR were considered in Refs.~\cite{AlaNod15,Cha15,MasLam21,AlaFal22,Pan22}.
The evolution of the relic neutrinos spin in stochastic gravitational fields was studied in Ref.~\cite{BayPen21}. The recent studies of the fermion spin evolution in external gravitational fields were reviewed in Ref.~\cite{Ver22}.

Using the quasiclassical approach, here, we study neutrino spin oscillations
in the particle scattering off a rotating BH. In this
situation, neutrinos are in the flat spacetime asymptotically. Hence,
we can control their `in' and `out' spin states. The gravitational scattering
of fermions, including neutrinos, was studied in Refs.~\cite{Lam05,DolDorLas06,Sor12}.
We analyzed this problem in Refs.~\cite{Dvo20a,Dvo20,Dvo21} accounting
for only the equatorial neutrino motion.

Now, for the first time, we discuss this problem in a quite general
form. The incoming flux of neutrinos is parallel to the equatorial
plane. However we do not restrict ourselves to the equatorial motion.
Particles can propagate both above and below the equatorial plane.
We take into account the change of the neutrino latitude in the scattering.
Moreover, the realistic magnetic field is accounted for in our work.
The neutrino interaction with a magnetic field is owing to the nonzero
neutrino magnetic moment~\cite{Giu16}. We suppose that a neutrino
is a Dirac particle.

The motivation for this work was the direct observation of the event
horizon silhouette of the supermassive BHs (SMBHs) in the centers of M87~\cite{Aki19}
and our Galaxy~\cite{Aki22}. These observations are the first direct
tests of GR in the strong field limit. The images
in Refs.~\cite{Aki19,Aki22} are formed by photons emitted by the
hot gas in the accretion disks around these SMBHs~\cite{DokNaz20}. The review of the analytical studies of BH shadows is provided in Ref.~\cite{PerTsu22}.
Accretion disks in some active galactic nuclei can be hot and dense
enough to emit both photons, protons and secondary neutrinos~\cite{KimMurMes21}.
The observation of these objects in a neutrino telescope (see, e.g.,
Ref.~\cite{Aar18}) should account for the neutrino spin precession
in strong external fields including gravity. Moreover, we can imagine
a hypothetical situation when a flux of neutrinos, e.g., from a core-collapsing
supernova (SN), is gravitationally lensed by BH~\cite{EirRom08}.

This work is organized in the following way. First, in Sec.~\ref{sec:EQUATIONS},
we formulate the main equations for the description of the general
motion of ultrarelativistic neutrinos in the Kerr metric, as well as
the spin evolution equation in the curved spacetime under the influence
of an electromagnetic field. Then, we fix the parameters of the system
and represent electromagnetic and gravi-electromagnetic fields in
the chosen geometry of the spacetime in Sec.~\ref{sec:DESCR}. In
Sec.~\ref{sec:RES}, we find the measurable fluxes of scattered neutrinos
accounting for their spin precession in the given external fields.
Finally, in Sec.~\ref{sec:DISC}, we summarize and discuss the possibility
to detect the predicted effects for SN neutrinos. The evolution of the helicity of an ultrarelativistic neutrino in its scattering in an arbitrary gravitational field is studied in Appendix~\ref{sec:HELTHEOR}.

\section{Motion of a neutrino in Kerr metric and the particle spin evolution\label{sec:EQUATIONS}}

The spacetime outside a rotating BH is described by the Kerr metric.
In Boyer-Lindquist coordinates $x^{\mu}=(t,r,\theta,\phi)$, this
metric has the form,
\begin{equation}\label{eq:Kerrmetr}
  \mathrm{d}s^{2} = g_{\mu\nu}\mathrm{d}x^{\mu}\mathrm{d}x^{\nu} =
  \left(
    1-\frac{rr_{g}}{\Sigma}
  \right)
  \mathrm{d}t^{2}+2\frac{rr_{g}a\sin^{2}\theta}{\Sigma}\mathrm{d}t\mathrm{d}\phi-\frac{\Sigma}{\Delta}\mathrm{d}r^{2}-
  \Sigma\mathrm{d}\theta^{2}-\frac{\Xi}{\Sigma}\sin^{2}\theta\mathrm{d}\phi^{2},
\end{equation}
where
\begin{equation}
  \Delta=r^{2}-rr_{g}+a^{2},
  \quad
  \Sigma=r^{2}+a^{2}\cos^{2}\theta,
  \quad
  \Xi=
  \left(
    r^{2}+a^{2}
  \right)
  \Sigma+rr_{g}a^{2}\sin^{2}\theta.
\end{equation}
Here we use units where the gravitational constant is equal to one. In
this situation, the mass of BH is $M=r_{g}/2$ and its angular momentum
is $J=Ma$, where $r_{g}$ is the Schwarzschild radius. The BH spin is directed upward with respect to the equatorial plane $\theta = \pi/2$.

A test particle in the Kerr metric has three integrals of motion:
the energy, $E$, the angular momentum, $L$, and the Carter constant,
$Q$. If we study the scattering problem, $Q>0$. The law of motion
of a test particle can be found in quadratures~\cite{Cha83},
\begin{equation}\label{eq:trtheta}
  t= E
  \left[
    \int\frac{r^{2}\mathrm{d}r}{\sqrt{R}}+a^{2}\int\frac{\mathrm{d}\theta}{\sqrt{\Theta}}\cos^{2}\theta
  \right]+
  r_{g}\int\frac{r\mathrm{d}r}{\sqrt{R}\Delta}[r^{2}E-a(L-aE)],
\end{equation}
where $R(r)=[(r^{2}+a^{2})E-aL]^{2}-\Delta[Q+(L-aE)^{2}]$ and $\Theta(\theta)=Q+\cos^{2}\theta\left[a^{2}E^{2}-L^{2}/\sin^{2}\theta\right]$. Here we consider an ultrarelativistic neutrino. The form of a trajectory
can be also determined in quadratures~\cite{Cha83},
\begin{equation}\label{eq:phirtheta}
  \phi=a\int\frac{\mathrm{d}r}{\sqrt{R}\Delta}[(r^{2}+a^{2})E-aL]+\int\frac{\mathrm{d}\theta}{\sqrt{\Theta}}
  \left(
    \frac{L}{\sin^{2}\theta}-aE
  \right),
\end{equation}
and
\begin{equation}\label{eq:rtheta}
  \int\frac{\mathrm{d}r}{\sqrt{R}}=\pm\int\frac{\mathrm{d}\theta}{\sqrt{\Theta}}.
\end{equation}
One should choose the sign in Eq.~(\ref{eq:rtheta}), e.g., for an
incoming particle at $r\to\infty$ and then keep the choice for the
whole trajectory. The signs in radial integrals in Eqs.~(\ref{eq:trtheta})
and~(\ref{eq:phirtheta}) depend whether a neutrino approaches or
moves away from BH. The description of the motion of a test particle
near a Kerr BH can be made in two ways. One can either solve the system
of the geodesics equations as in Ref.~\cite{Zak91}. Alternatively,
we can analyze the integrals in Eqs.~(\ref{eq:trtheta})-(\ref{eq:rtheta}).

The scattering of a test particle has three main difficulties. First,
some neutrinos in the incoming beam can fall into BH. Hence, we should
take only the specific values of $Q$ and $L$ for incoming neutrinos
(see, e.g., Ref.~\cite{GraLupStr18}). Second, a neutrino can make
multiple revolutions around BH, i.e. the polar angle $|\phi|$ in
Eq.~(\ref{eq:phirtheta}) can be greater than $2\pi$. Third, the
latitude of an incoming neutrino does not coincide with that for an
outgoing particle. Moreover, the $\theta$-dependence in Eq.~(\ref{eq:rtheta})
can be oscillating. One should account for these facts in the analysis
of the particle trajectories.

Now, we can describe the general dynamics of the neutrino spin in the Kerr metric. Following Refs.~\cite{Dvo06,PomKhr98}, we define the invariant neutrino
spin $\bm{\zeta}$ in rest frame in the locally Minkowskian coordinates
$x_{a}=e_{a}^{\ \mu}x_{\mu}$, where
\begin{align}\label{eq:vierbKerr}
  e_{0}^{\ \mu}= &
  \left(
    \sqrt{\frac{\Xi}{\Sigma\Delta}},0,0,\frac{arr_{g}}{\sqrt{\Delta\Sigma\Xi}}
  \right),
  \quad
  e_{1}^{\ \mu}=
  \left(
    0,\sqrt{\frac{\Delta}{\Sigma}},0,0
  \right),
  \nonumber
  \\
  e_{2}^{\ \mu}= &
  \left(
    0,0,\frac{1}{\sqrt{\Sigma}},0
  \right),
  \quad
  e_{3}^{\ \mu}=
  \left(
    0,0,0,\frac{1}{\sin\theta}\sqrt{\frac{\Sigma}{\Xi}}
  \right),
\end{align}
are the vierbein vectors, which satisfy the relation $e_{a}^{\ \mu}e_{b}^{\ \nu}g_{\mu\nu}=\eta_{ab}$, where $\eta_{ab}=\text{diag}(1,-1,-1,-1)$ is the Minkowski metric tensor.

In our problem, we consider the neutrino gravitational scattering
off a rotating BH surrounded by a thin magnetized accretion disk.
In this situation, only the interaction with gravity and with the
poloidal component of the magnetic field contribute to the neutrino
spin evolution (see Sec.~\ref{sec:DESCR} below). The vector $\bm{\zeta}$ obeys the equation
\begin{equation}\label{eq:nuspinrot}
  \frac{\mathrm{d}\bm{\bm{\zeta}}}{\mathrm{d}t}=2(\bm{\bm{\zeta}}\times\bm{\bm{\Omega}}),
\end{equation}
where
\begin{equation}\label{eq:vectG}
  \bm{\bm{\Omega}}=\frac{1}{U^{t}}
  \left\{
    \frac{1}{2}
    \left[
      \mathbf{b}_{g}+\frac{1}{1+u^{0}}
      \left(
        \mathbf{e}_{g}\times\mathbf{u}
      \right)
    \right]+
    \mu
    \left[
      u^{0}\mathbf{b}-\frac{\mathbf{u}(\mathbf{u}\mathbf{b})}{1+u^{0}}+(\mathbf{e}\times\mathbf{u})
    \right]
  \right\}.
\end{equation}
Here $u^{a}=(u^{0},\mathbf{u})=e_{\ \mu}^{a}U^{\mu}$, $U^{\mu}=(U^{t},U^{r},U^{\theta},U^{\phi})$
is the four velocity of a neutrino in the world coordinates, $\mathbf{e}_{g}$
and $\mathbf{b}_{g}$ are the components of the tensor $G_{ab}=(\mathbf{e}_{g},\mathbf{b}_{g})=\gamma_{abc}u^{c}$,
$\gamma_{abc}=\eta_{ad}e_{\ \mu;\nu}^{d}e_{b}^{\ \mu}e_{c}^{\ \nu}$
are the Ricci rotation coefficients, the semicolon stays for the covariant
derivative, $f_{ab}=e_{a}^{\ \mu}e_{b}^{\ \nu}F_{\mu\nu}=(\mathbf{e},\mathbf{b})$
is the electromagnetic field tensor in the locally Minkowskian frame,
with $F_{\mu\nu}=\partial_{\mu}A_{\nu}-\partial_{\nu}A_{\mu}$ being
the electromagnetic field in the world coordinates. We suppose that a
neutrino is a Dirac particle having the magnetic moment $\mu$. The
details of the derivation of Eqs.~(\ref{eq:nuspinrot}) and~(\ref{eq:vectG})
can be found in Ref.~\cite{Dvo13}.

\section{Parameters of the system\label{sec:DESCR}}

We consider SMBH with the mass $M=10^{8}M_{\odot}$ surrounded by
a thin magnetized accretion disk. The incoming flux of neutrinos is
taken to be parallel to the equatorial plane of BH. The consideration
of a thin disk makes it possible to neglect the effect of the neutrino
electroweak interaction with matter (see, e.g., Ref.~\cite{DvoStu02})
of a disk since only a small fraction of neutrinos moves in the equatorial
plane. We mentioned in Sec.~\ref{sec:EQUATIONS} that the latitude
of a neutrino can oscillate in its motion, i.e. a particle can cross
the equatorial plane multiple times. However, the path inside a thin
disk for such neutrinos is short. Hence, we neglect the electroweak
interaction with plasma of a disk. Moreover, the consideration of
a thick disk makes the problem more complex. Indeed, a plasma cannot
rotate on circular orbits in such a disk. Slim accretion disks were
mentioned in Ref.~\cite{Abr88} to be possible around SMBHs.

The plasma rotation in a disk generates the magnetic field. Both poloidal
and toroidal components are created. However, if the disk is thin,
we can neglect the toriodal component since it is inside a disk,
even despite the toroidal field can be quite strong. The poloidal field
is taken to correspond to the following vector potential in the world
coordinates~\cite{Wal74}:
\begin{equation}\label{eq:Atphi}
  A_{t}=aB
  \left[
    1-\frac{rr_{g}}{2\Sigma}(1+\cos^{2}\theta)
  \right],
  \quad
  A_{\phi}=-\frac{B}{2}
  \left[
    r^{2}+a^{2}-\frac{a^{2}rr_{g}}{\Sigma}(1+\cos^{2}\theta)
  \right]
  \sin^{2}\theta,
\end{equation}
where $B$ is the magnetic field strength, which is uniform and is
along the BH spin at $r\to\infty$.

However, the assumption that $B$ is finite at $r\to\infty$ is unphysical
since the magnetic field is created by the plasma motion in a disk,
which has the finite size. Thus, we should suppose that $B\to0$ at
$r\to\infty$. For example, we can take that $B\propto B_{0}r^{-5/4}$~\cite{BlaPay82},
where $B_{0}$ is the strength of the magnetic field in the vicinity
of BH at $r\sim r_{g}$. We suppose that $B_{0}=10^{-2}B_{\mathrm{Edd}}=3.2\times10^{2}\,\text{G}$
for $M=10^{8}M_{\odot}$, where $B_{\mathrm{Edd}}$ is the Eddington
limit for the magnetic field which arrests the accretion~\cite{Bes10}.

A Dirac neutrino is taken to have the magnetic moment in the range
$\mu=(10^{-14}-10^{-13})\mu_{\mathrm{B}}$, where $\mu_{\mathrm{B}}$
is the Bohr magneton. The smaller value of $\mu=10^{-14}\mu_{\mathrm{B}}$
is consistent with the model independent upper bound on the Dirac
neutrino magnetic moment established in Ref.~\cite{Bel05}. The greater $\mu=10^{-13}\mu_{\mathrm{B}}$
considered is below the best astrophysical
upper limit on the neutrino magnetic moment in Ref.~\cite{Via13}.

We consider the incoming flux of neutrinos moving from the point with the coordinates
$(r, \theta, \phi)_s= (\infty,\pi/2,0)$. In this case, the asymptotic neutrino velocity is $\mathbf{u}_{\pm\infty}=(\pm|u_{1}|,0,0)$,
i.e. incoming and outgoing neutrinos move oppositely and along the
first axis in the locally Minkowskian frame. Instead of Eq.~(\ref{eq:nuspinrot}),
we can deal with an effective Schr\"{o}dinger equation $\mathrm{i}\dot{\psi}=H\psi$,
where $H=-\mathcal{U}_{2}(\bm{\bm{\sigma}}\cdot\bm{\bm{\Omega}})\mathcal{U}_{2}^{\dagger}$,
$\mathcal{U}_{2}=\exp(\mathrm{i}\pi\sigma_{2}/4)$, $\bm{\sigma}=(\sigma_{1},\sigma_{2},\sigma_{3})$
are the Pauli matrices, and $\bm{\bm{\Omega}}$ is given in Eq.~(\ref{eq:vectG}).
We suppose that initially all neutrinos are left polarized, i.e. their
initial helicity is $h_{-\infty}=(\bm{\zeta}_{-\infty}\mathbf{u}_{-\infty})/|\mathbf{u}_{-\infty}|=-1$.
It corresponds to the initial effective wavefunction $\psi_{-\infty}^{\mathrm{T}}=(1,0)$.
We are interested in the survival probability $P_{\mathrm{LL}}$,
which indicates how many neutrinos remain left polarized after the
scattering. If the wavefunction of outgoing neutrinos is $\psi_{+\infty}^{\mathrm{T}}=(\psi_{+\infty,1},\psi_{+\infty,2})$, then 
$P_{\mathrm{LL}}=|\psi_{+\infty,2}|^{2}$.

It is convenient to introduce the dimensionless variables, $r=xr_{g}$,
$L=yr_{g}E$, $a=zr_{g}$, $Q=wr_{g}^{2}E^{2}$. The components of
gravi-electromagnetic field $(\tilde{\mathbf{e}}_{g},\tilde{\mathbf{b}}_{g})=\tfrac{\mathrm{d}t}{\mathrm{d}r}(\mathbf{e}_{g},\mathbf{b}_{g})$
in Eq.~(\ref{eq:vectG}) have the form,
\begin{align}\label{eq:egbg}
  \tilde{e}_{g1} = & \frac{1}
  {2\sqrt{z^{2}\cos^{2}\theta(x^{2}-x+z^{2})+z^{2}x^{2}+z^{2}x+x^{4}}(x^{2}+z^{2}\cos^{2}\theta)^{2}}
  \notag
  \\
  & \times
  \Big\{
    r_{g}\frac{\mathrm{d}\phi}{\mathrm{d}r}
    \left[
      z^{3}\cos^{4}\theta(z^{2}-x^{2})-z\cos^{2}\theta(z^{4}+3x^{4})+z^{3}x^{2}+3zx^{4}
    \right]
    \notag
    \\
    & +
    \frac{\mathrm{d}t}{\mathrm{d}r}
    \left[
      z^{2}\cos^{2}\theta(z^{2}+x^{2})-z^{2}x^{2}-x^{4}
    \right]
  \Big\},
  \nonumber
  \displaybreak[2]
  \\
  \tilde{e}_{g2} = & \frac{z^{2}x\sin2\theta\sqrt{x^{2}-x+z^{2}}
  \left[
    \frac{\mathrm{d}t}{\mathrm{d}r}-r_{g}\frac{\mathrm{d}\phi}{\mathrm{d}r}z\sin^{2}\theta
  \right]}
  {2\sqrt{z^{2}\cos^{2}\theta(x^{2}-x+z^{2})+z^{2}x^{2}+z^{2}x+x^{4}}(x^{2}+z^{2}\cos^{2}\theta)^{2}},
  \nonumber
  \displaybreak[2]
  \\
  \tilde{e}_{g3} = & -\frac{
  \left[
    z^{2}x\sin2\theta r_{g}\frac{\mathrm{d}\theta}{\mathrm{d}r}(x^{2}-x+z^{2})+z^{2}\cos^{2}\theta(z^{2}-x^{2})-z^{2}x^{2}-3x^{4}
  \right]}
  {2
  \left[
      z^{4}\cos^{4}\theta(x^{2}-x+z^{2})+xz^{2}\cos^{2}\theta(2z^{2}x+2x^{3}-x^{2}+z^{2})+x^{6}+z^{2}x^{4}+z^{2}x^{3}
  \right]}
  \notag
  \\
  & \times
  \frac{z\sin\theta}{\sqrt{x^{2}-x+z^{2}}},
  \nonumber
  \displaybreak[2]
  \\
  \tilde{b}_{g1} = & \frac{\cos\theta}
  {\sqrt{z^{2}\cos^{2}\theta(x^{2}-x+z^{2})+z^{2}x^{2}+z^{2}x+x^{4}}(x^{2}+z^{2}\cos^{2}\theta)^{2}}
  \notag
  \\
  & \times
  \Big\{
    r_{g}
    \frac{\mathrm{d}\phi}{\mathrm{d}r}
    \left[
      z^{4}\cos^{4}\theta(x^{2}-x+z^{2})+2x^{2}z^{2}\cos^{2}\theta(x^{2}-x+z^{2})+z^{4}x+z^{2}x^{4}+2z^{2}x^{3}+x^{6}
    \right]
    \notag
    \\
    & -
    zx\frac{\mathrm{d}t}{\mathrm{d}r}(z^{2}+x^{2})
  \Big\},
  \nonumber
  \displaybreak[2]
  \\
  \tilde{b}_{g2} = & -\frac{\sin\theta\sqrt{x^{2}-x+z^{2}}}
  {2\sqrt{z^{2}\cos^{2}\theta(x^{2}-x+z^{2})+z^{2}x^{2}+z^{2}x+x^{4}}(x^{2}+z^{2}\cos^{2}\theta)^{2}}
  \notag
  \\
  & \times
  \Big\{
    r_{g}\frac{\mathrm{d}\phi}{\mathrm{d}r}
    \left[
      z^{4}\cos^{4}\theta(2x-1)+z^{2}\cos^{2}\theta(z^{2}+4x^{3}+x^{2})+2x^{5}-z^{2}x^{2}
    \right]
    \notag
    \\
    & +
    z\frac{\mathrm{d}t}{\mathrm{d}r}(x^{2}-z^{2}\cos^{2}\theta)
  \Big\},
  \nonumber
  \\
  \tilde{b}_{g3} = & \frac{z^{2}\sin\theta\cos\theta+r_{g}\frac{\mathrm{d}\theta}{\mathrm{d}r}x(x^{2}-x+z^{2})}
  {\sqrt{x^{2}-x+z^{2}}(x^{2}+z^{2}\cos^{2}\theta)},
\end{align}
where the derivatives with respect to $r$,
can be obtained on the basis of Eqs.~(\ref{eq:trtheta})-(\ref{eq:rtheta}). Despite the expressions for $\tilde{\mathbf{e}}_{g}$ and $\tilde{\mathbf{b}}_{g}$ in Eq.~\eqref{eq:egbg} are valid for particles with arbitrary masses, we consider mainly massless neutrinos moving along null geodesics.
Analogously we find the electromagnetic field $(\mathbf{e},\mathbf{b})$
in the locally Minkowskian frame,
\begin{align}\label{eq:eb}
  e_{1} = & \frac{\mu Bz
  \left[
    z^{2}\cos^{4}\theta(z^{2}-x^{2})+\cos^{2}\theta(z^{4}+2z^{2}x^{2}-3x^{4})-z^{2}x^{2}+x^{4}
  \right]}
  {2\sqrt{z^{2}\cos^{2}\theta(x^{2}-x+z^{2})+z^{2}x^{2}+z^{2}x+x^{4}}(x^{2}+z^{2}\cos^{2}\theta)^{2}},
  \nonumber
  \\
  e_{2} = & \frac{\mu Bxz^{3}\sin2\theta\sqrt{x^{2}-x+z^{2}}(1+\cos^{2}\theta)}
  {2\sqrt{z^{2}\cos^{2}\theta(x^{2}-x+z^{2})+z^{2}x^{2}+z^{2}x+x^{4}}(x^{2}+z^{2}\cos^{2}\theta)^{2}},
  \nonumber
  \\
  b_{1} = & \frac{\mu B\cos\theta
  \left[
    (x^{2}+z^{2}\cos^{2}\theta)^{2}(x^{2}-x+z^{2})+x(x^{4}-z^{4})
  \right]}
  {\sqrt{z^{2}\cos^{2}\theta(x^{2}-x+z^{2})+z^{2}x^{2}+z^{2}x+x^{4}}(x^{2}+z^{2}\cos^{2}\theta)^{2}},
  \nonumber
  \\
  b_{2} = & \mu B\sin\theta\sqrt{x^{2}-x+z^{2}}
  \notag
  \\
  & \times
  \frac{
  \left[
    z^{4}\cos^{4}\theta(1-2x)+z^{2}\cos^{2}\theta(z^{2}-x^{2}-4x^{3})-z^{2}x^{2}-2x^{5}
  \right]}
  {2\sqrt{z^{2}\cos^{2}\theta(x^{2}-x+z^{2})+z^{2}x^{2}+z^{2}x+x^{4}}(x^{2}+z^{2}\cos^{2}\theta)^{2}},
\end{align}
and $e_{3}=b_{3}=0$. These fields correspond to the vector potential
in Eq.~(\ref{eq:Atphi}).

It should be noted that Eq.~(\ref{eq:vectG}) was derived for a massive
particle. However, it has the finite limit for a massless neutrino.
For example, the components of the vector $\tilde{\mathbf{u}}=\tfrac{\mathbf{u}}{1+u^{0}}$
are
\begin{align}\label{eq:v}
  \tilde{u}_{1} = & \frac{\sqrt{z^{2}\cos^{2}\theta(x^{2}-x+z^{2})+z^{2}x^{2}+z^{2}x+x^{4}}}{z^{2}+x^{2}-x}
  \frac{\mathrm{d}r}{\mathrm{d}t},
  \nonumber
  \\
  \tilde{u}_{2}= & \frac{\sqrt{z^{2}\cos^{2}\theta(x^{2}-x+z^{2})+z^{2}x^{2}+z^{2}x+x^{4}}}{\sqrt{z^{2}+x^{2}-x}}r_{g}
  \frac{\mathrm{d}\theta}{\mathrm{d}t},
  \nonumber
  \\
  \tilde{u}_{3}= & \frac{\sin\theta}{(x^{2}+z^{2}\cos^{2}\theta)\sqrt{(x^{2}-x+z^{2})}}
  \notag
  \\
  & \times
  \left\{
    r_{g}\frac{\mathrm{d}\phi}{\mathrm{d}t}\left[z^{2}\cos^{2}\theta(z^{2}+x^{2}-x)+z^{2}x^{2}+z^{2}x+x^{4}\right]-zx
  \right\} ,
\end{align}
where the time derivatives can be obtained again using Eqs.~(\ref{eq:trtheta})-(\ref{eq:rtheta}).
The ratio
\begin{equation}
  \frac{u^{0}}{U^{t}}=
  \frac{\sqrt{z^{2}\cos^{2}\theta+x^{2}}\sqrt{z^{2}+x^{2}-x}}
  {\sqrt{z^{2}\cos^{2}\theta(x^{2}-x+z^{2})+z^{2}x^{2}+z^{2}x+x^{4}}},
\end{equation}
is also finite.

Finally, we rewrite the effective Schr\"{o}dinger equation in the form,
\begin{equation}\label{eq:Schrodx}
  \mathrm{i}\frac{\mathrm{d}\psi}{\mathrm{d}x}=H_{x}\psi,\quad H_{x}=r_{g}H\frac{\mathrm{d}t}{\mathrm{d}r}.
\end{equation}
Equations~(\ref{eq:trtheta})-(\ref{eq:rtheta}) and~(\ref{eq:Schrodx})
completely define the evolution of spinning ultrarelativistic neutrinos
in their scattering off a rotating BH.

\section{Results\label{sec:RES}}

Before we discuss the results, some description of the computational
details should be present since they are not so trivial. The initial
beam of neutrinos has a circular form with the radius $\sim10r_{g}$.
We form it on the distance $r_{m}=10^{2}r_{g}$ from BH. This beam
is taken to be denser towards its center to probe trajectories close
to the BH surface. Initially it has $3.2\times10^{3}$ neutrinos.
After eliminating particles which fall to BH, we deal with $\gtrsim10^{3}$
neutrinos. The partition of a trajectory from $r_{m}$ to the turn
point is made with $4\times10^{3}$ nodes.

To reconstruct the trajectory, we start with Eq.~(\ref{eq:rtheta}).
The $r$-integral is computed numerically, whereas the $\theta$-integral
is expressed in terms of the elliptic integrals. Then, we find the
number of extrema for the latitude and $\theta$ for each point of
the trajectory using the Jacobi elliptic functions. Having the coordinates
$(r,\theta)$ in any point of the trajectory, we compute $\phi$ using
Eq.~(\ref{eq:phirtheta}) and express the angular integral again
in terms of the elliptic integrals. In principle, the spin evolution
does not depend on $\phi$, as one can see in Eqs.~(\ref{eq:egbg})-(\ref{eq:v}).
Nevertheless, the $\phi$-dependence of the trajectory is required
for the representation of the results. Eventually, we obtain the angles
$\theta_{\mathrm{obs}}$ and $\phi_{\mathrm{obs}}$ which correspond
to an outgoing neutrino. The adopted calculation procedure guarantees
that $0<\theta_{\mathrm{obs}}<\pi$. However, as we mentioned above,
a particle can make multiple revolutions around BH. Thus, we have
to project $\phi_{\mathrm{obs}}$ to the interval $(0,2\pi)$. 

One situates an observer in the point with the coordinates $(r,\theta,\phi)_o = (\infty,\theta_{\mathrm{obs}},\phi_{\mathrm{obs}})$. In our work, we vary both $0<\theta_{\mathrm{obs}}<\pi$ and $0<\phi_{\mathrm{obs}}<2\pi$, whereas the coordinates of the source of the neutrino beam are fixed, $(r,\theta,\phi)_s = (\infty,\pi/2,0)$. Thus, e.g., the point with $\theta_{\mathrm{obs}}=\pi/2$ and $\phi_{\mathrm{obs}}=\pi$ corresponds to the forward neutrino scattering, and that with $\theta_{\mathrm{obs}}=\pi/2$ and $\phi_{\mathrm{obs}}=0$ (or $2\pi$) to the backward one. Any pair of the final algular coordinates $(\theta_{\mathrm{obs}},\phi_{\mathrm{obs}})$ correspond to the specific $L$ and $Q$ in the incoming beam.

It should be noted that we have to reconstruct the whole trajectory
since we are interested in the neutrino spin evolution rather than
in the calculation of a differential cross section for scalar particles.
Thus we have to build the trajectory from $r_{m}$ to the turn point
for incoming particles and from the turn point to $r_{m}$ for outgoing
ones. Since a rotating BH does not correspond a central field, the reconstruction
of both branches of the trajectory consumes computational resources.
The details for the finding of a trajectory in a general gravitational
scattering of an ultrarelativistic spinless particle can be found,
e.g., in Ref.~\cite{Boz08}.

Then, we solve Eq.~(\ref{eq:Schrodx}) along the trajectory. Since
the dependence $\theta(r)$ is known only in certain nodes, we cannot
use a precise Runge-Kutta solver with an adaptive stepsize. Instead,
we apply the Euler method to integrate Eq.~(\ref{eq:Schrodx}). It
significantly reduces the accuracy of computations. Analogously to
the reconstruction of the trajectory, Eq.~(\ref{eq:Schrodx}) is
solved separately for both branches of the trajectory. We use the
result at the turn point as the initial condition for the second branch
of the trajectory. Finally, we find $P_{\mathrm{LL}}$ and plot it
for any $(\theta_{\mathrm{obs}},\phi_{\mathrm{obs}})$ point. We use the
2D cubic interpolation to get the smooth surface which is represented
as a contour plot.

In some cases, contour plots have white gaps meaning the insufficient
number of neutrinos scattered to these areas. It happens especially
for a rapidly rotating BH. The gaps can be eliminated by increasing
initial number of particles. However, in this case the computational
time increases significantly. Since we have the limited access to
the computer facilities, this problem will be tackled in a future work.

Standard model neutrinos are created as left polarized particles.
If their spin is flipped because of the interaction with an external
field, we observe the effective reduction of the neutrino flux since
a detector can also register left neutrinos only. By definition, the
flux of particles, gravitationally scattered in a certain solid angle
$\mathrm{d}\varOmega$, is proportional to the differential cross-section,
$F\propto\mathrm{d}\sigma/\mathrm{d}\varOmega$. Thus, if we deal
with spinning neutrinos, the observed flux in the wake of their scattering
is $F_{\nu}=P_{\mathrm{LL}}F_{0}$, where $F_{0}$ is the flux of
scalar particles. It is the consequence of the fact that, in the quasiclassical approximation, used in our work (see also Ref.~\cite{PomKhr98}), the  spin of a particle does not influence its motion.

The flux of scalar ultrarelativistic particles, $F_{0}$, corresponds to the situation when only their propagation along null geodesics is accounted for.
The very detailed study of $F_{0}$, which includes not only ultrarelativistic particles, is provided in Ref.~\cite{Gru14}. Our goal is to study the ratio $P_{\mathrm{LL}} = F_{\nu}/F_{0}$ for neutrinos
gravitationally scattered off a rotating BH accounting for the neutrino
interaction with a magnetic field in an accretion disk. If we have the map of $P_{\mathrm{LL}}(\phi_{\mathrm{obs}},\theta_{\mathrm{obs}})$ for all scattered particles, we can reconstruct the flux of spinning neutrinos by combining our results with those in Ref.~\cite{Gru14}.

First, we turn off the magnetic field and consider only the contribution
of gravity to the neutrino spin-flip. We recall that we deal with ultrarelativistic
neutrinos. The statement that ultrarelativistic fermions conserve
their polarization is valid in flat spacetime. In curved spacetime,
it may be not the case. For example, it was claimed
in Refs.~\cite{Mer95,SinMobPap04} that a massless neutrino can change its polarization in the gravitational scattering. Earlier, we established that
the polarization of ultrarelativistic neutrinos is conserved in their
gravitational scattering off non-rotating~\cite{Dvo20} and rotating~\cite{Dvo21}
BHs. However, those results were obtained only for the neutrino motion
in the equatorial plane.

\begin{figure}
  \centering
  \subfigure[]
  {\label{fig:grava}
  \includegraphics[scale=.35]{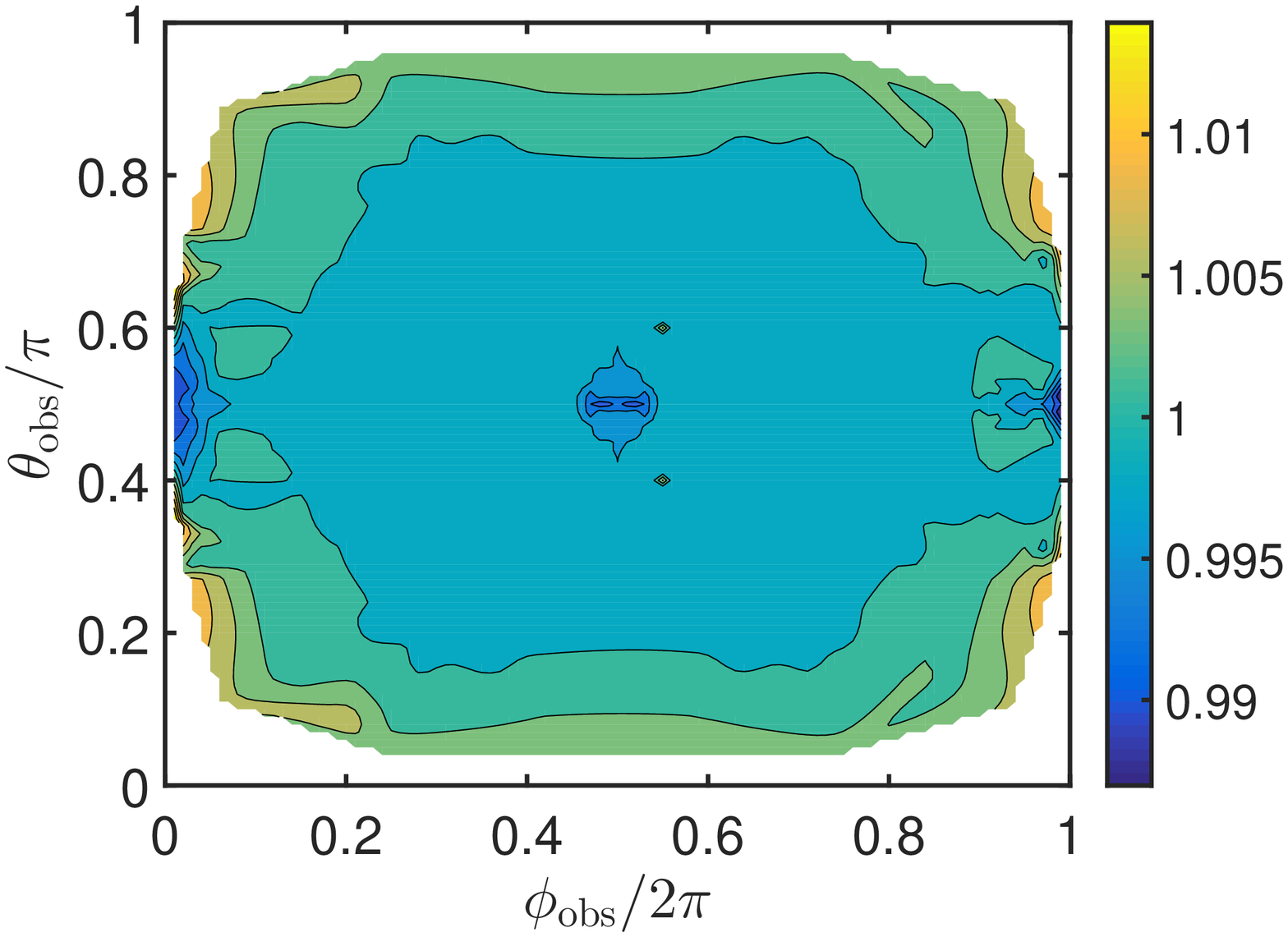}}
  \hskip-.3cm
  \subfigure[]
  {\label{fig:gravb}
  \includegraphics[scale=.35]{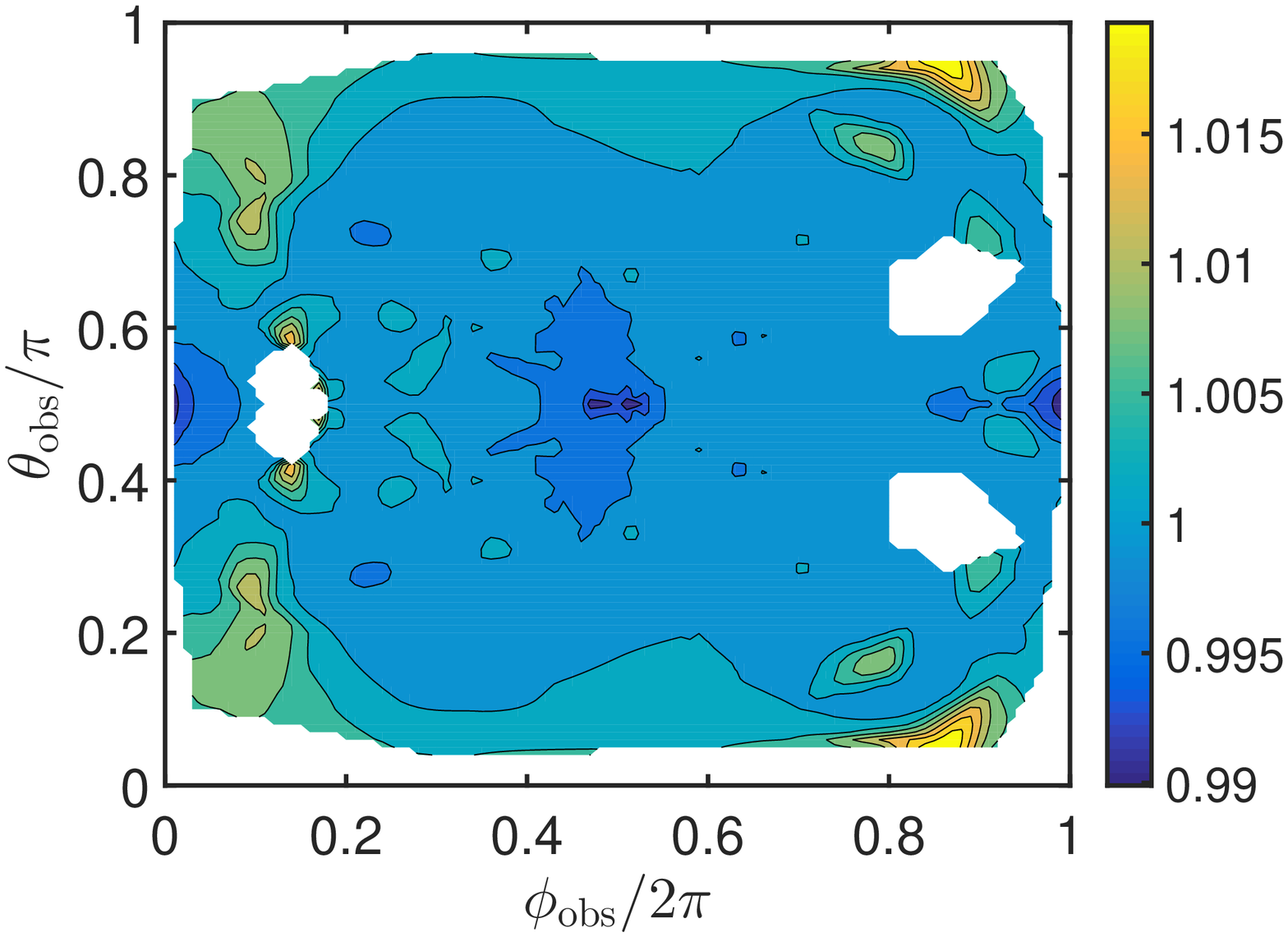}}
  \protect
\caption{The ratio of the fluxes of spinning neutrinos $F_{\nu}$ and the scalar
particles $F_{0}$ gravitationally scattered off BHs with different
angular momenta. (a) $z=10^{-2}$ ($a=2\times10^{-2}M$); (b) $z=0.49$
($a=0.98M$).\label{fig:grav}}
\end{figure}

To examine the issue of the neutrino polarization for arbitrary trajectories,
we plot $F_{\nu}/F_{0}$ for BHs with different spins in Fig.~\ref{fig:grav}.
We present the cases of an almost non-rotating BH with $a=2\times10^{-2}M$
in Fig.~\ref{fig:grava} and an almost maximally rotating BH with
$a=0.98M$ in Fig.~\ref{fig:gravb}. We can see that, in both situations,
$P_{\mathrm{LL}}\approx1$ with the accuracy $\sim1\%$. Thus, we get that the gravitational interaction only does not lead to the spin-flip
of ultrarelativistic neutrinos. This result generalizes our findings
in Refs.~\cite{Dvo20,Dvo21}. In Appendix~\ref{sec:HELTHEOR}, we prove the general statement that the helicity of an ultrarelativistic particle is constant when it scatters in an arbitrary gravitational field.

\begin{figure}
  \centering
  \subfigure[]
    {\label{fig:magna}
    \includegraphics[scale=.35]{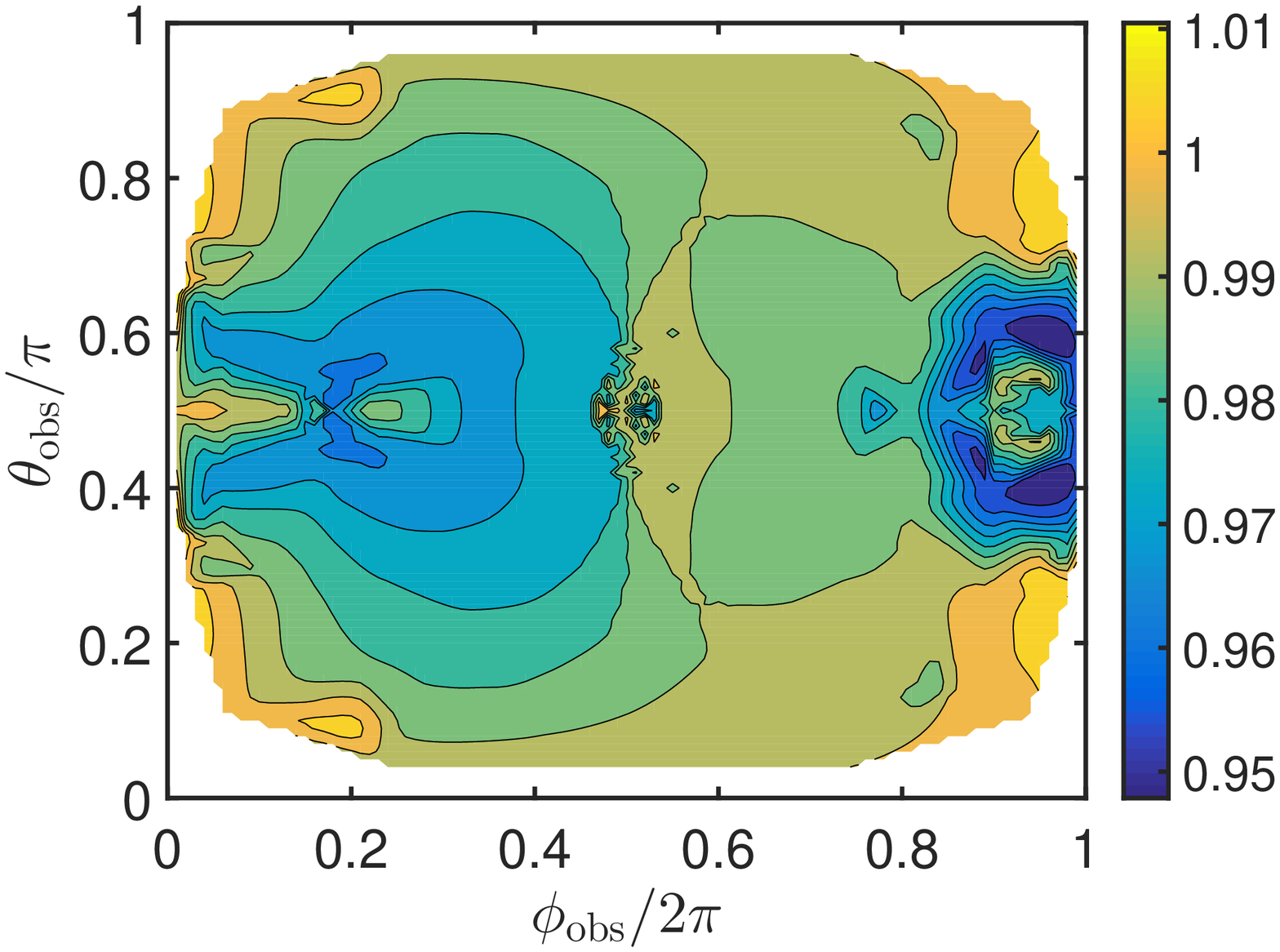}}
  \hskip-.3cm
  \subfigure[]
    {\label{fig:magnb}
    \includegraphics[scale=.35]{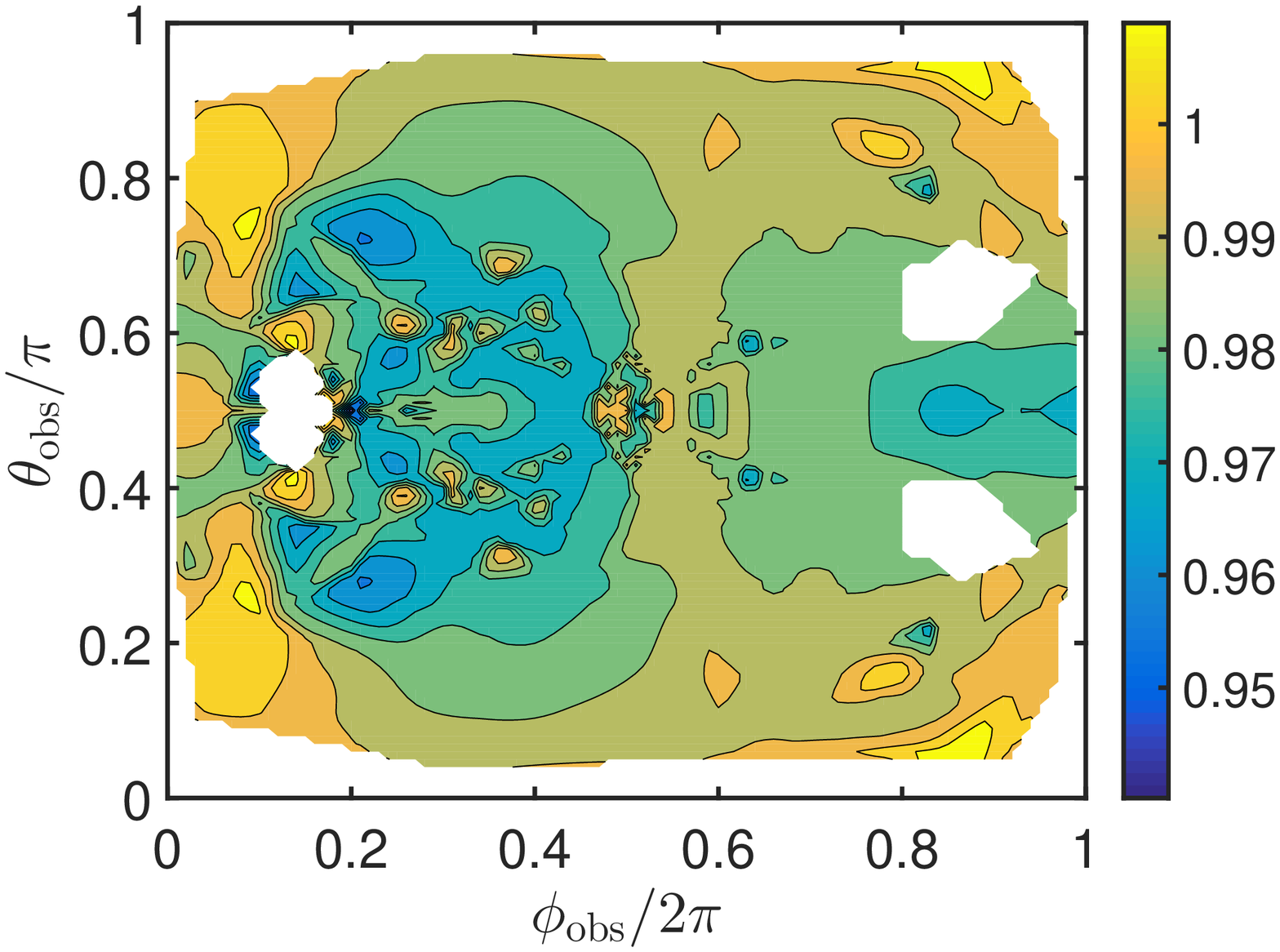}}
  \\
  \subfigure[]
    {\label{fig:magnc}
    \includegraphics[scale=.35]{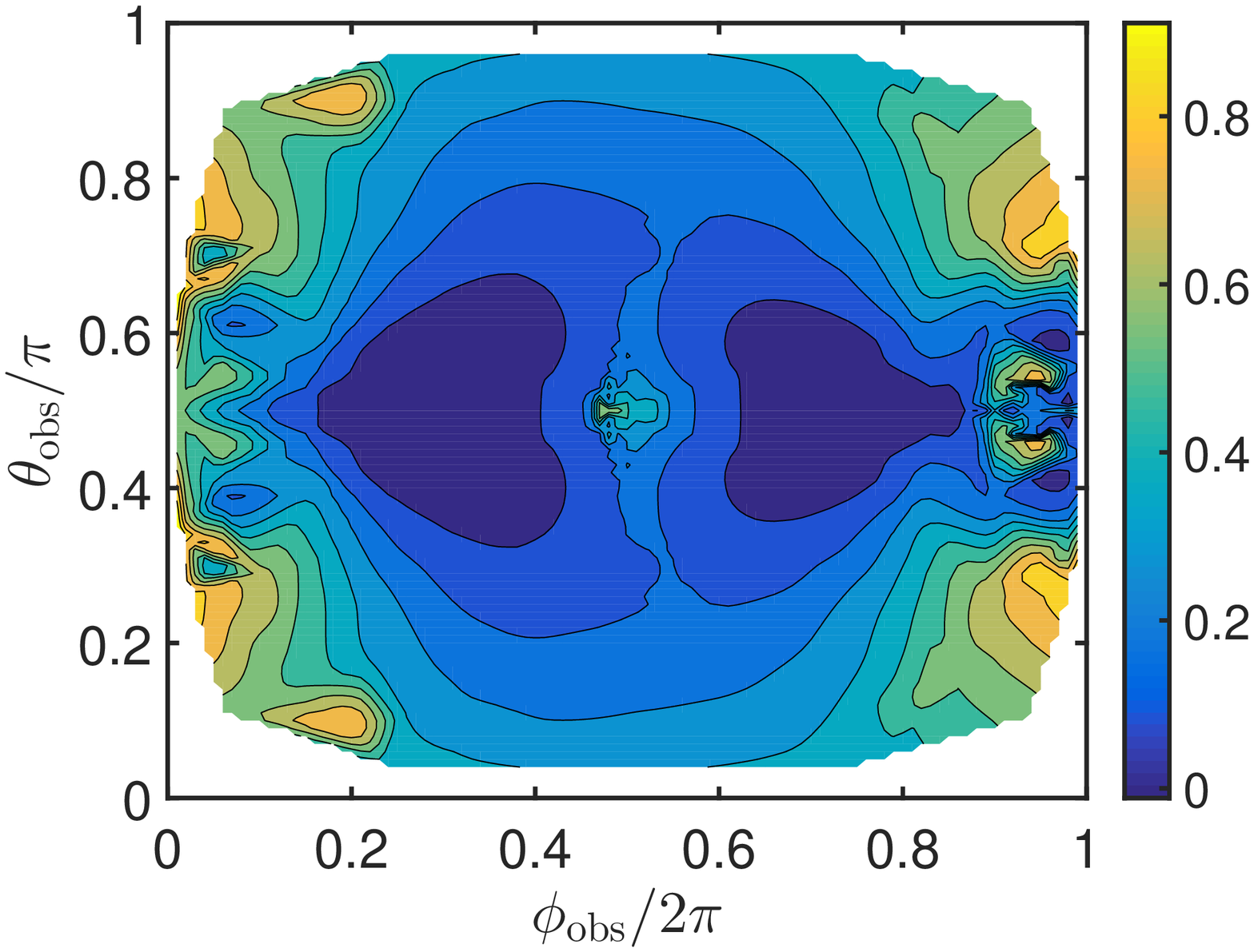}}
  \hskip-.3cm
  \subfigure[]
    {\label{fig:magnd}
    \includegraphics[scale=.35]{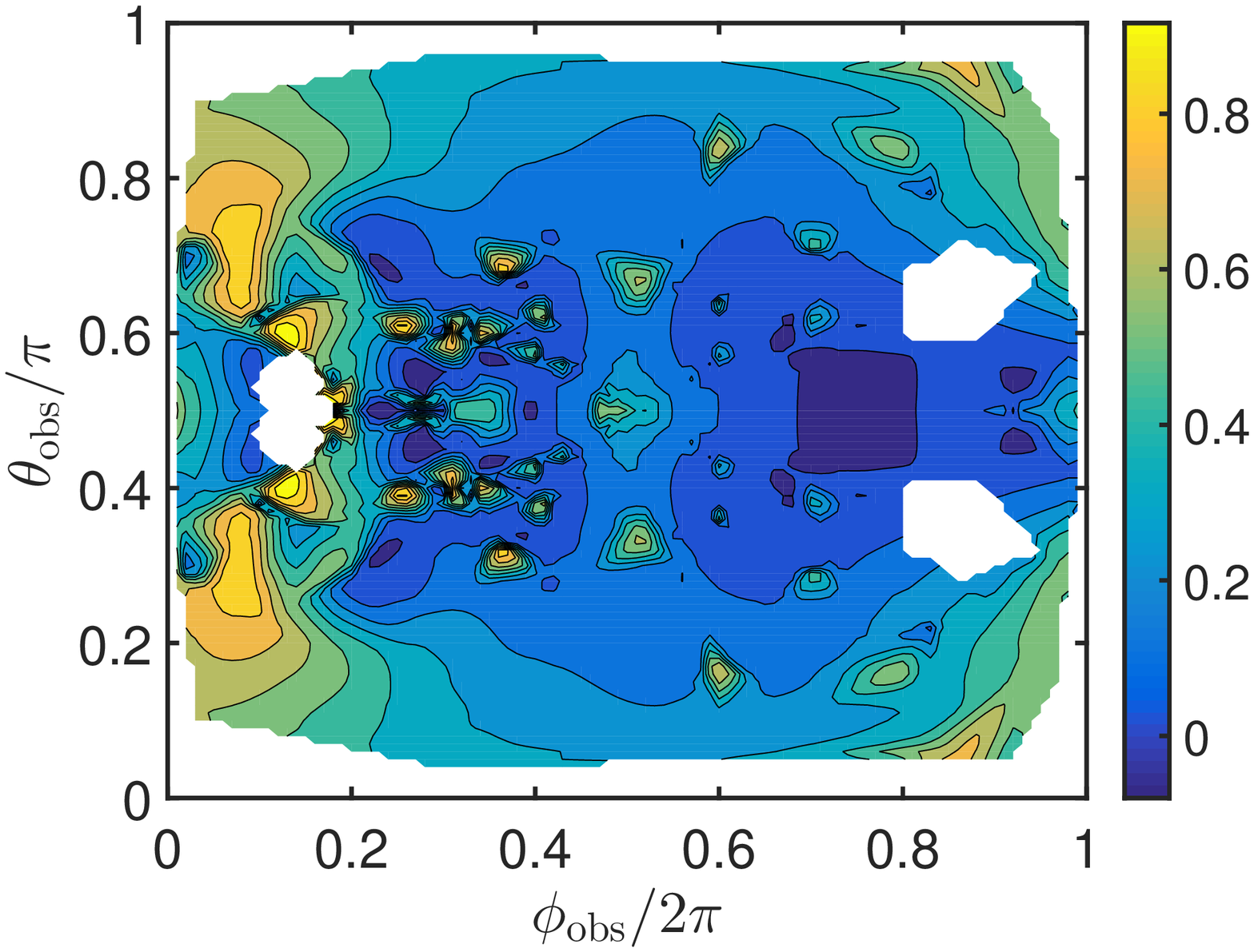}}
  \protect
\caption{The same as in Fig.~\ref{fig:grav} accounting for the neutrino magnetic
interaction. Panels (a) and (b): $V_{\mathrm{B}}=2.7\times10^{-2}$;
panels (c) and (d): $V_{\mathrm{B}}=2.7\times10^{-1}$. Panels (a)
and (c): $z=10^{-2}$ ($a=2\times10^{-2}M$); panels (b) and (d):
$z=0.49$ ($a=0.98M$).\label{fig:magn}}
\end{figure}

Now we can account for the neutrino interaction with the magnetic
field in an accretion disk. In this case, $F_{\nu}/F_{0}$ is depicted
in Fig.~\ref{fig:magn} for the different BH spins and the different
values of the magnetic parameter $V_{\mathrm{B}}=\mu B_{0}r_{g}$.
If we consider SMBH with $M=10^{8}M_{\odot}$, fix the magnetic field
$B_{0}=3.2\times10^{2}\,\text{G}$, and vary the neutrino magnetic
moment in the range $10^{-14}\mu_{\mathrm{B}}<\mu<10^{-13}\mu_{\mathrm{B}}$
(see Sec.~\ref{sec:DESCR}), we get that $2.7\times10^{-2}<V_{\mathrm{B}}<2.7\times10^{-1}$.
As in Fig.~\ref{fig:grav}, we consider two cases. Figures~\ref{fig:magna}
and~\ref{fig:magnc} correspond to $a=2\times10^{-2}M$, whereas
Figs.~\ref{fig:magnb} and~\ref{fig:magnd} to $a=0.98M$.

Taking the very conservative value of $\mu=10^{-14}\mu_{\mathrm{B}}$,
we can see in Figs.~\ref{fig:magna} and~\ref{fig:magnb} that
the flux of neutrinos is reduced by up to $(5\pm1)\%$ compared to
that of scalar particles. This result is in agreement with Ref.~\cite{Dvo21},
where the similar reduction factor was obtained while considering
the equatorial neutrino motion. However, if we increase the neutrino
magnetic moment by one order of magnitude to $\mu=10^{-13}\mu_{\mathrm{B}}$,
we can observe in Figs.~\ref{fig:magnc} and~\ref{fig:magnd}
that the neutrino flux is almost suppressed in certain directions.
We mentioned in Sec.~\ref{sec:DESCR} that such neutrino magnetic
moments are not excluded by the astrophysical observations.

\begin{figure}
  \centering
  \subfigure[]
    {\label{fig:unitara}
    \includegraphics[scale=.35]{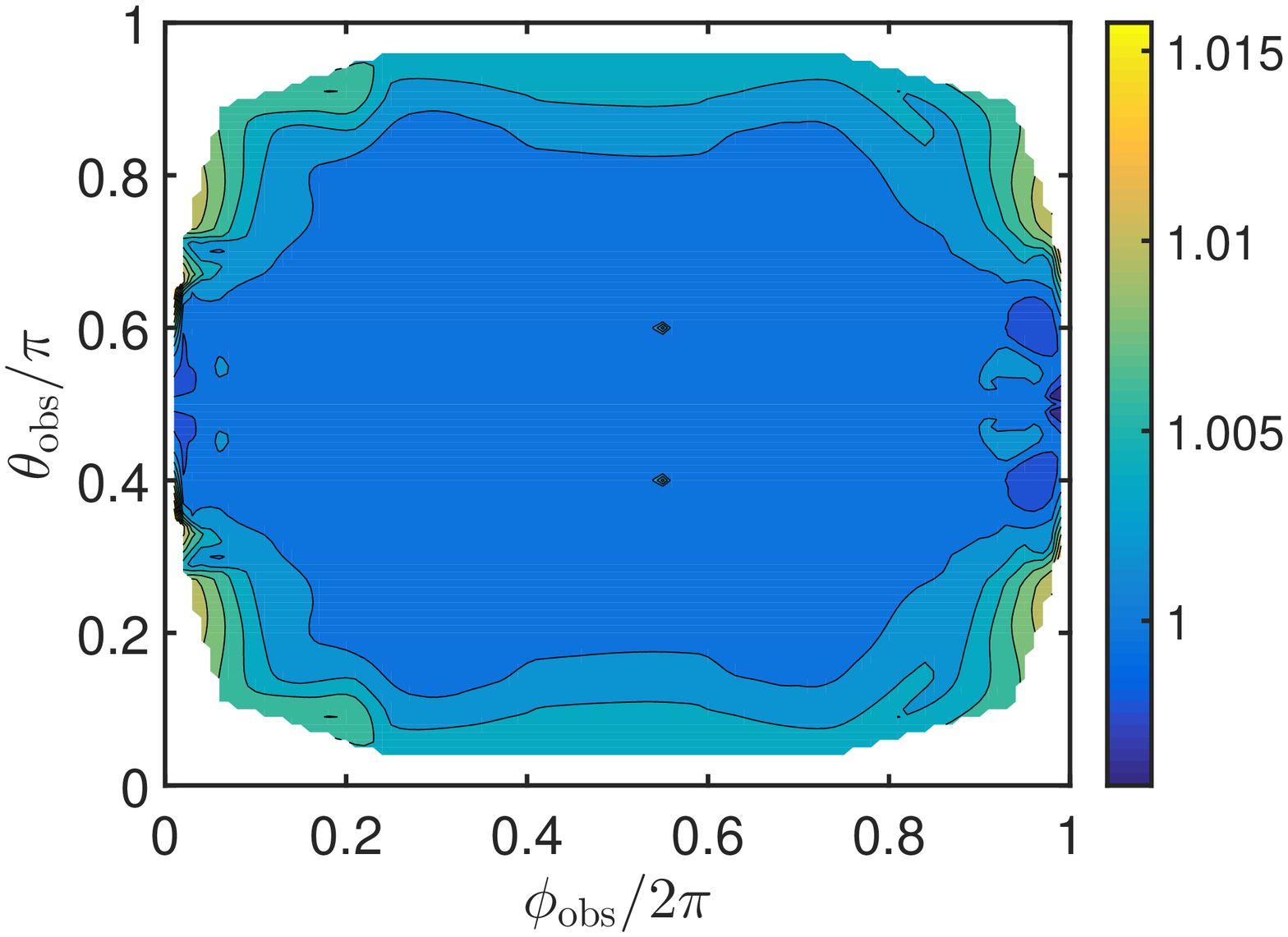}}
  \hskip-.3cm
  \subfigure[]
    {\label{fig:unitarb}
    \includegraphics[scale=.35]{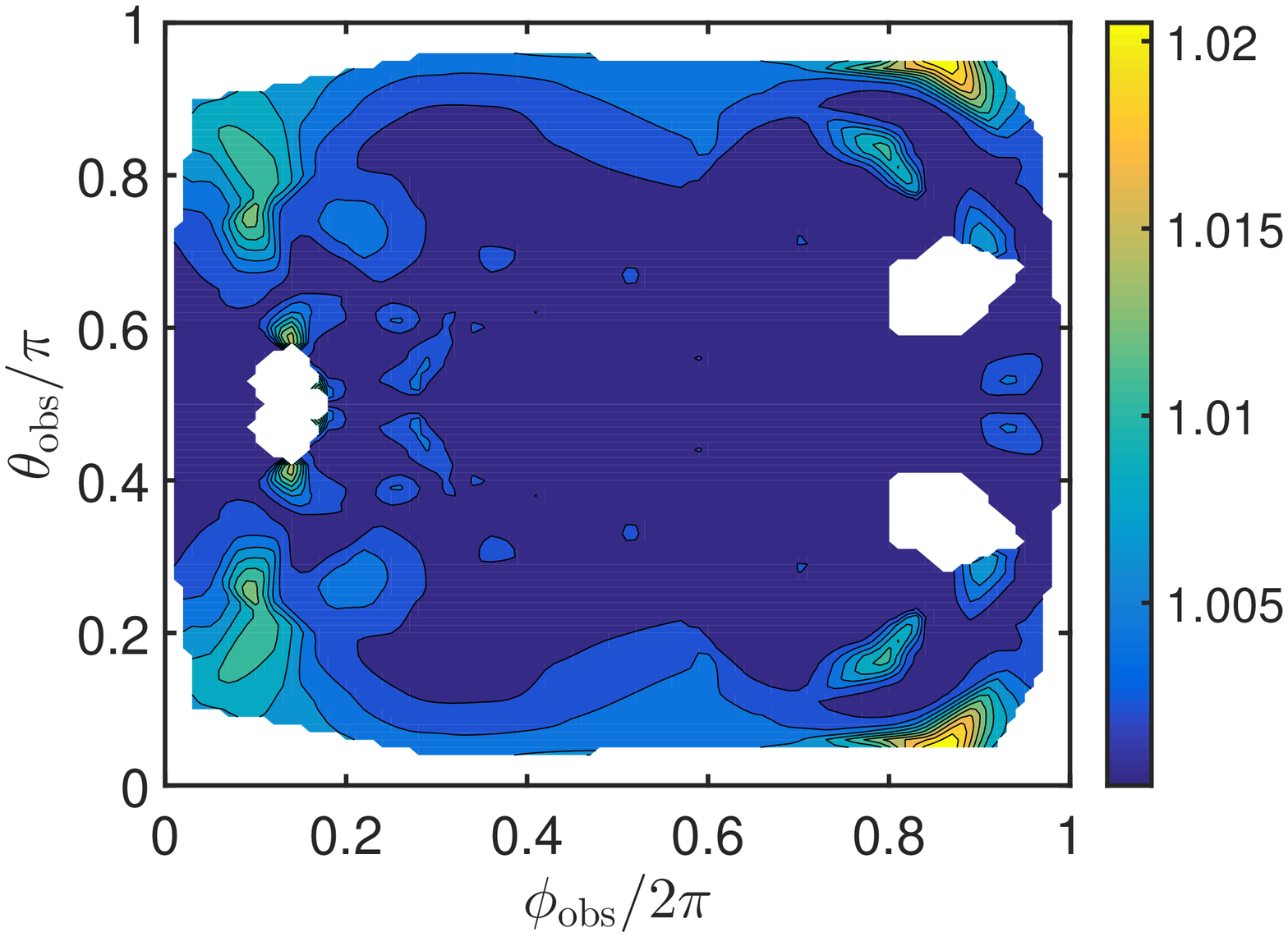}}
  \\
  \subfigure[]
    {\label{fig:unitarc}
    \includegraphics[scale=.35]{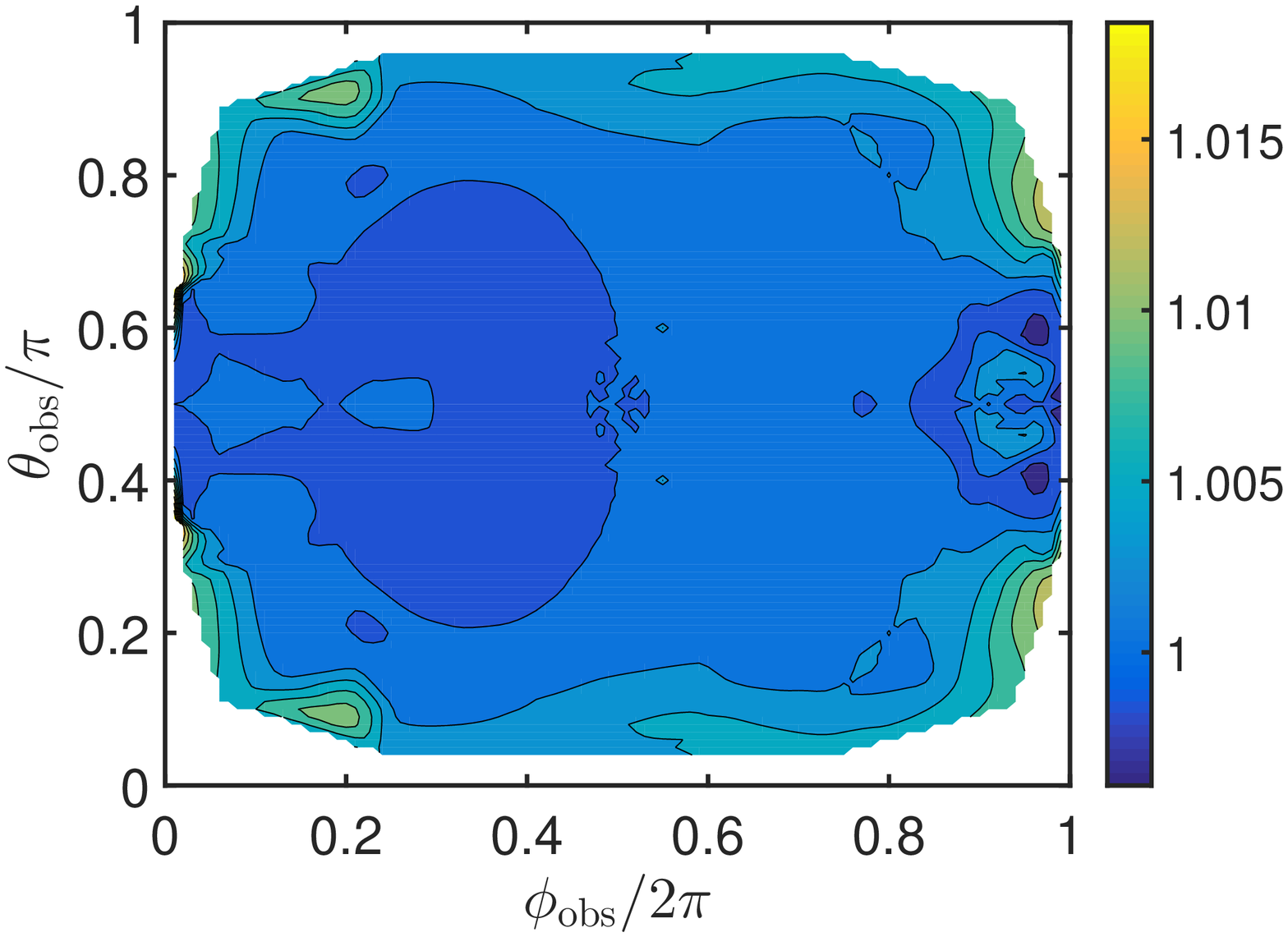}}
  \hskip-.3cm
  \subfigure[]
    {\label{fig:unitard}
    \includegraphics[scale=.35]{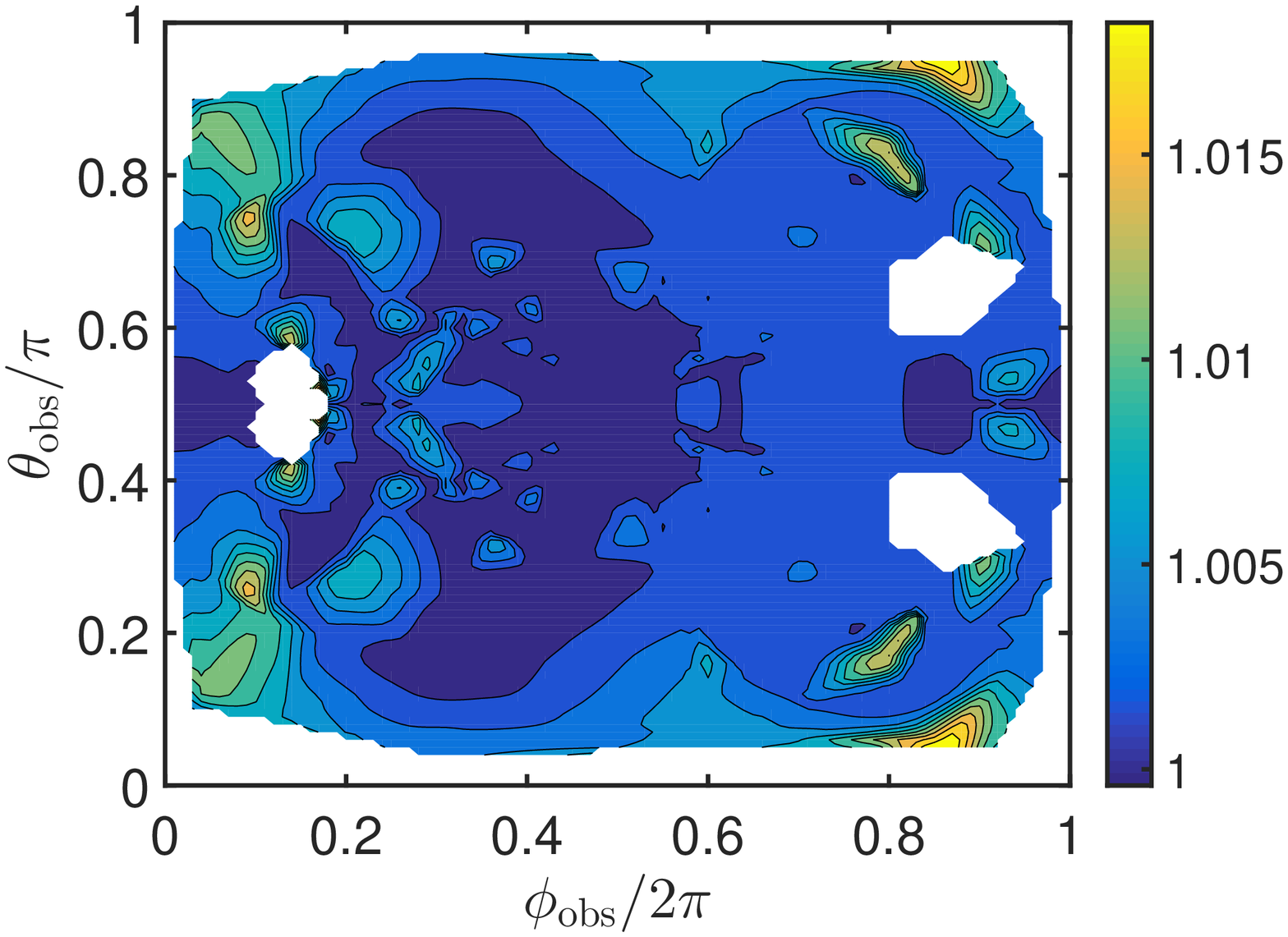}}
  \protect
\caption{The sum $P_\mathrm{LL} + P_\mathrm{LR}$ for neutrinos gravitationally scattered off BHs with different angular momenta accounting for the neutrino magnetic
interaction. The panels here correspond to these in Fig.~\ref{fig:magn}. Namely, panels (a) and (b): $V_{\mathrm{B}}=2.7\times10^{-2}$;
panels (c) and (d): $V_{\mathrm{B}}=2.7\times10^{-1}$. Panels (a)
and (c): $z=10^{-2}$ ($a=2\times10^{-2}M$); panels (b) and (d):
$z=0.49$ ($a=0.98M$).\label{fig:unitar}}
\end{figure}

To check the conservation of the probability in our simulations, in Fig.~\ref{fig:unitar}, we plot the quantity $P_\mathrm{LL} + P_\mathrm{LR}$ for neutrinos scattered off BH with a magnetized accretion disk. The transition probability $P_\mathrm{LR}$ is computed directly basing on the solution of Eq.~\eqref{eq:Schrodx} as $P_\mathrm{LR} = |\psi_{+\infty,1}|^{2}$. One can see in Fig.~\ref{fig:unitar} that the unitarity condition $P_\mathrm{LL} + P_\mathrm{LR} = 1$ is fulfilled for any neutrino trajectory with the accuracy $\sim 1\%$. The validity of the same condition can be checked for the purely gravitational scattering shown in Fig.~\ref{fig:grav}. It means that our simulations are reliable.

Finally, we notice that the plots in Figs.~\ref{fig:grav} and~\ref{fig:magn}
are symmetric with respect to equator $\theta_{\mathrm{obs}}=\pi/2$.
It is the consequence of the symmetry of the metric in Eq.~(\ref{eq:Kerrmetr})
to the reflection $\theta\to\pi-\theta$.

\section{Discussion\label{sec:DISC}}

In the present work, we have studied spin effects in the neutrino
gravitational scattering off BH. Particles were supposed to move on
general trajectories unlike previous works~\cite{Dvo20,Dvo21}, where
only the motion in the equatorial plane was considered. The effects
of gravity were accounted for exactly in the reconstruction of the
neutrino trajectory, i.e. we have considered the strong gravitational
lensing. The neutrino spin evolution was studied in the locally Minkowskian
frame by solving the effective Schr\"{o}dinger equation. Neutrinos were
supposed to be left polarized before scattering. If the neutrino helicity
changes, we would observe the effective reduction of the outgoing
neutrino flux.

We have found that the gravitational interaction only does not result
in the change of the neutrino polarization. The flux of outgoing spinning
neutrinos is seen in Fig.~\ref{fig:grav} to be identical to that
of scalar particles. It generalizes our findings in Refs.~\cite{Dvo20,Dvo21},
where we studied neutrinos moving in the equatorial plane only. This
result also corrects the claims in Refs.~\cite{Mer95,SinMobPap04}
that gravity can change the helicity of an ultrarelativistic fermion. The general theorem that the helicity of an ultrarelativistic neutrino remains constant in the particle scattering by an arbitrary gravitational field has been proven in Appendix~\ref{sec:HELTHEOR}.

To produce the neutrino spin-flip we have added the neutrino interaction
with a poloidal magnetic field which is generated in an accretion
disk surrounding BH. We have assumed that the disk is slim. Thus one
takes into account neither matter effects nor a toroidal magnetic
field for the neutrino spin evolution. Considering the strength of
the magnetic field which is allowed in realistic SMBHs and the moderate
value of the Dirac neutrino magnetic moment $\mu=10^{-14}\mu_{\mathrm{B}}$,
we have obtained that the observed neutrino flux can be reduced by
$\sim5\%$ in certain directions; cf. Figs.~\ref{fig:magna} and~\ref{fig:magnb}.
This result is in agreement with Ref.~\cite{Dvo21}. If we take greater
magnetic moment $\mu=10^{-13}\mu_{\mathrm{B}}$, that still does not
violate astrophysical constrains, the neutrino flux turns out to be
vanishing for some scattering directions; see Figs.~\ref{fig:magnc}
and~\ref{fig:magnd}. These our results are valid for both non-rotating, with $a\ll M$,
and maximally rotating, with $a\lesssim M$, BHs. For example, SMBH in the center of M87 has a quite great $a\approx 0.9$~\cite{Tam19}.

\begin{figure}
\centering
\includegraphics[scale=0.3]{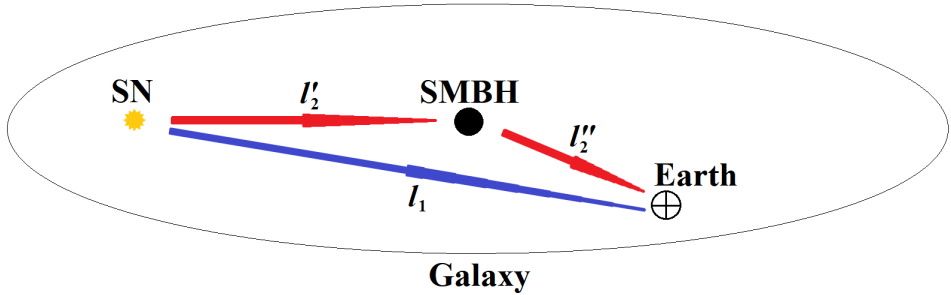}
\caption{The schematic illustration of the neutrino gravitational scattering off SMBH, which is depicted by the
black blob, in the center of our Galaxy. Neutrinos are emitted in
a core-collapsing SN shown by the yellow asterisk. Then, they travel
along the paths 1 and 2. The path 1, represented by the blue arrow, goes directly to the Earth depicted
by the symbol~$\oplus$. The path 2, shown by red arrows, accounts for the gravitational
lensing of neutrinos by SMBH. The distances $l_1$, $l_2'$, and $l_2''$ are between SN and the Earth, SN and SMBH, SMBH and the Earth, respectively.\label{fig:schem}}
\end{figure}

Described neutrino spin oscillations in the particle gravitational
scattering can be potentially observed if a core-collapsing SN, which
emits huge amount of neutrinos, explodes in our Galaxy. The event
is schematically depicted in Fig.~\ref{fig:schem}. Suppose that
SN explodes somewhere in the Galaxy. Using current or future neutrino
telescopes, we detect the direct flux of neutrinos $F_1$ along the path 1
shown in Fig.~\ref{fig:schem} by the blue arrow. Then, one starts to look for a neutrino signal in the direction to the galactic
center where SMBH is situated. These particles propagate along the path 2 depicted in Fig.~\ref{fig:schem} by the red arrows.
Such neutrinos are gravitationally lensed and their polarizations
are affected by the magnetic field in the vicinity of SMBH. The flux $F_2$ is related to $F_1$ by $F_2 = \tfrac{l_1^2}{4\pi l_2^{\prime 2}l_2^{\prime\prime 2}}\left( \tfrac{\mathrm{d}\sigma}{\mathrm{d}\varOmega} \right) F_1$, where $l_1$, $l_2'$, and $l_2''$ are the distances between objects in Fig.~\ref{fig:schem}. The differential cross-section $\mathrm{d}\sigma/\mathrm{d}\varOmega$ corresponds to the scattering angles fixed by the positions of the objects.

One has that $\mathrm{d}\sigma/\mathrm{d}\varOmega = r_g^2 f$, where $r_g^2 = 1.4\times 10^{24}\,\text{cm}^2$ for the SMBH in Sgr A$^*$ and the function $f(\theta_\text{obs},\phi_\text{obs})$ has great values for forward and backward scatterings (see, e.g., Refs.~\cite{ColDelWil73,Dvo20}), as well as in caustics~\cite{RauBla94,Boz08}. If $l_1 \sim \ l_2' \sim l_2'' \sim 10\,\text{kpc}$, then $F_2 = 8.2\times 10^{-23} F_1 f$. The observed flux of neutrinos, if a core-collapsing SN takes place in our Galaxy,
is estimated by $F_1 \sim7\times10^{3}$ events for the JUNO detector~\cite{An16}
and $F_1 \sim7\times10^{4}$ events for the Hyper-Kamiokande detector~\cite{Abe18}.

To get the sizable flux of lensed neutrinos, $F_2$, these particles should be observed, e.g., very close to the SMBH surface. Thus, SN, SMBH and the Earth should be on one line practically, with $\theta_{\mathrm{obs}}\approx\pi/2$ and $\phi_{\mathrm{obs}}\approx\pi$. If this case, the function $f$ can become great enough to exceed the factor $8.2\times 10^{-23} F_1 = 5.7\times 10^{-18}$ for the Hyper-Kamiokande detector. Using Fig.~\ref{fig:magnc} or Fig.~\ref{fig:magnd}, we obtain that the observed neutrino flux can be $50\%$ reduced because of spin oscillations provided that the neutrino magnetic interaction is strong enough.


\appendix

\section{Conservation of helicity in the gravitational scattering of utrarelativistic neutrinos\label{sec:HELTHEOR}}

In this appendix, we examine the evolution of the helicity of utrarelativistic neutrinos in their scattering in an arbitrary gravitational field. We prove that the helicity is conserved.

The general quasiclassical spin evolution in a gravitational field was studied in Refs.~\cite{PomKhr98,Dvo06}. The four vector of a fermion spin $s^a$, defined in a locally Minkowskian frame, evolves as
\begin{equation}\label{eq:saevol}
  \frac{\mathrm{d}s^{a}}{\mathrm{d}t}=\frac{1}{U^t}G^{ab}s_{b},
\end{equation}
where $U^t$ and $G_{ab}$ are given in Sec.~\ref{sec:EQUATIONS}. Besides Eq.~\eqref{eq:saevol}, we should take into account the evolution of the particle velocity $u^a$ which has the form,
\begin{equation}\label{eq:uaevol}
  \frac{\mathrm{d}u^{a}}{\mathrm{d}t}=\frac{1}{U^t}G^{ab}u_{b}.
\end{equation}
We rewrite Eqs.~\eqref{eq:saevol} and~\eqref{eq:uaevol} in the three dimensional form,
\begin{align}\label{eq:dot3D}
  \frac{\mathrm{d}\bm{\zeta}}{\mathrm{d}t}= & \frac{1}{U^{t}}
  \left[
    \bm{\zeta}\times
    \left\{
      \mathbf{b}_{g}+\frac{1}{1+u^{0}}(\mathbf{e}_{g}\times\mathbf{u})
    \right\}  
  \right],
  \notag
  \\
  \frac{\mathrm{d}\mathbf{u}}{\mathrm{d}t} = & \frac{1}{U^{t}}(\mathbf{u}\times\mathbf{b}_{g}),
  \quad
  \frac{\mathrm{d}u^{0}}{\mathrm{d}t} = \frac{1}{U^{t}}(\mathbf{u}\cdot\mathbf{e}_{g}),
\end{align}  
where the vectors $\bm{\zeta}$, $\mathbf{e}_{g}$, and $\mathbf{b}_{g}$ are also given in Sec.~\ref{sec:EQUATIONS}.

If a neutrino is ultrarelativistic, then both $u^{0} \to \infty$ and $\mathbf{u}\to\infty$, whereas the velocity in the locally Minkowskian frame $\mathbf{v}=\mathbf{u}/u^{0}$ is a unit vector, $|\mathbf{v}|\to 1$. Thus, the helicity of such neutrinos is $h=(\bm{\zeta}\cdot\mathbf{v})$. Its time evolution, measured by a distant observer, reads
\begin{equation}\label{eq:doth}
  \frac{\mathrm{d}h}{\mathrm{d}t}=(\dot{\bm{\zeta}}\cdot\mathbf{v})+(\bm{\zeta}\cdot\dot{\mathbf{v}})\to
  -\frac{(\bm{\zeta}\cdot\mathbf{e}_{g})}{U^t}
\end{equation}
where we use Eq.~\eqref{eq:dot3D} and the fact that a neutrino is ultrarelativistic.

In the gravitational scattering, a neutrino propagates in the region outside the BH surface, i.e. $e_{a}^{\ \mu}$ in Eq.~\eqref{eq:vierbKerr} are nonzero and finite. In fact, in the case of the Kerr metric, the allowed values of $L$ and $Q$ are outside the BH shadow which is greater than the BH horizon. Thus, $U^t = e_{a}^{\ 0} u^a \to \infty$ for an ultrarelativistic neutrino. Therefore, using Eq.~\eqref{eq:doth}, we get that $h=\text{const}$.

It is important that we consider the scattering problem, where a source and a detector of neutrinos are in asymptotically flat spacetime, i.e. we measure the neutrino helicity with respect to the world time $t$. In this case, the neutrino helicity is conserved. In other situations, e.g., when the helicity of ultrarelativistic neutrinos is measured by a comoving observer, it can change in a gravitational field.

\section*{Acknowledgments}

I am thankful to A.~F.~Zakharov for the useful discussion.

\end{document}